\pdfoutput=1
\documentclass[10pt,aps,prb,amsmath,amssymb,twocolumn,letterpaper,%
         showpacs,balancelastpage,final,citeautoscript,floatfix]{revtex4-1}
\usepackage{graphicx}
\usepackage{enumerate}
\usepackage{microtype}
\usepackage[bookmarks=false,colorlinks]{hyperref}

\newcommand{\LNO }{La$_4$Ni$_3$O$_{10}$}



\begin{document}
\title{Crystal structure stability and electronic properties of layered nickelate La$_4$Ni$_3$O$_{10}$}
\author{Danilo Puggioni}
\affiliation{Department of Materials Science and Engineering, 
Northwestern University, Evanston, Illinois 60208, USA}

\author{James M.\ Rondinelli}
\affiliation{Department of Materials Science and Engineering, 
Northwestern University, Evanston, Illinois 60208, USA}

\begin{abstract}
We investigate the crystal structure and the electronic properties of the trilayer nickelate \LNO\  by means of quantum mechanical calculations in the framework of the density functional theory. We find that, at low temperature, \LNO\ undergoes a hitherto unreported structural phase transition and transforms to a new monoclinic $P2_1/a$ phase. 
This phase exhibits electronic properties in agreement with recent angle-resolved photoemission spectroscopy data reported in H.\ Li \emph{et al}., Nat.\ Commun.\ \textbf{8}, 704 (2017) and should be considered in models  focused on explaining the observed $\sim$140\,K  metal-to-metal phase transition.
\end{abstract}

%\pacs{%
%75.85.+t, %multiferroics
%71.45.Gm, %collective effects in correlation
%77.80.B$-$, %phase transitions, FE
%71.20.$-$b,%band structure
%}
\date{\today}

\maketitle

%\sloppy
%\section{Introduction}
\emph{Introduction}.---%
Since the discovery of high-temperature superconductivity in the layered hole-doped cuprate La$_{2-x}$Ba$_x$CuO$_4$\cite{Bednorz/Muller:1986}, numerous explorations have been performed to find copper-free transition metal oxides with comparable or higher transition temperatures ($T_c$).\cite{heatingup:2017} %
Superconductivity has been found in layered ruthenates\cite{Maeno_et_al:1994}  and cobaltates\cite{takada:2003}; however, the superconductivity in these $t_{2g}$-electron systems differs from that in the cuprates and $T_c$ is rather low. 
The band structure of high-$T_c$ superconducting cuprates (HTSC) derives from the $3d^9$ electronic configuration of Cu$^{2+}$ with one hole in the $d_{x^2-y^2}$ orbital. 
Layered nickelates are $e_g$-electron compounds and broadly exhibit an electronic configuration similar to that of cuprates, which makes them ideal candidates to search for high-$T_c$ superconductivity. 

In addition, the  Ruddlesden-Popper (RP) nickelates $R_4$Ni$_3$O$_{10}$ ($R=$ La, Pr, and Nd) 
share the same crystal habit as those of the HTSCs. The most common oxidation states are Ni$^{3+}$ and Ni$^{2+}$, however, which results in electronic properties that are quite different from cuprates.\cite{Anisimov/Bukhvalov/Rice:1999}  
Nonetheless, layered nickelates exhibit a number of intriguing physical properties and phase transitions.\cite{PhysRevMaterials.1.021801,PhysRevLett.118.177601,PhysRevB.67.014511}
For example, trilayered perovskite $R_4$Ni$_3$O$_{10}$ compounds are members of a small family of oxides that exhibit metallic conductivity even at low temperatures.\cite{ZHANG1995236}  
Interesting, these compounds undergo a metal-to-metal transition between 140~K and 165~K.\cite{ZHANG1995236,Tkalich:1991,Haoxiang:2017} Recently,  Li {\it et al.}\cite{Haoxiang:2017} analyzed the Fermi surface of the trilayer nickelate \LNO\  using angle-resolved photoemission spectroscopy (ARPES) and density functional theory (DFT) band structure calculations. 
The study describes the similarities and differences between the low-energy electronic structure of layered cuprates and nickelates. Moreover, the experimental evolution of the Fermi surface as a function of temperature shows the appearance of a pseudogap between 120 and 150~K, which suggests a connection with the anomaly observed in the temperature-dependent resistivity  $\sim$140~K.\cite{Tkalich:1991,Haoxiang:2017} This prior work did not consider the possibility of a structural transition, despite the established importance of lattice instabilities on both metallicity and superconductivity.\cite{PhysRevB.45.7650,PhysRevB.91.195115,motizuki2012structural}
Indeed, Li {\it et al.}\cite{Haoxiang:2017,priv_comm} interpret their data by assuming that the layered nickelate \LNO\ maintains its $Bmab$ symmetry over the entire temperature range analyzed.\cite{Tkalich:1991}

In this Rapid Communication, we show that the orthorhombic $Bmab$ phase of \LNO\ is unstable at low-temperature and undergoes a structural phase transition to monoclinic $P2_1/a$ using {\it ab initio} calculations. We find that the low-energy electronic structure of this monoclinic phase is in better agreement with previously published ARPES data\cite{Haoxiang:2017}, suggesting that the structural phase transition is a necessary component to achieving a complete  description of the physics in \LNO.

\emph{Methods}.---%
We perform first-principles density functional non-spin-polarized calculations with the local-density approximation (LDA)\cite{Perdew/Zunger:1981} as implemented in the 
Vienna Ab initio Simulation Package (VASP)\cite{Kresse/Furthmuller:1996b}  
with the projector augmented wave (PAW) method \cite{Blochl/Jepsen/Andersen:1994}
 to treat the core and valence electrons using the following electronic configurations: 
 5s$^2$5p$^6$6s$^2$5d$^1$ (La), 
 3d$^8$4s$^2$ (Ni), 
 2s$^2$2p$^4$ (O).
A $7\times7\times3$ Monkhorst-Pack $k$-point mesh\cite{Monkhorst/Pack:1976}  and a 600~eV planewave cut-off are used. For structural relaxations, we relax the atomic positions (forces to be less than 0.1~meV~\AA$^{-1}$) and use  Gaussian smearing (0.02~eV width) for the Brillouin zone (BZ) integrations. 
We use the AMPLIMODES software\cite{Orobengoa:ks5225, Perez-Mato:sh5107} for the group theoretical analyses.
For all data discussed, we find that the Perdew-Burke-Ernzerhof (PBE) generalized gradient approximation functional\cite{Perdew/Burke/Ernzerhof:1996} and the generalized gradient approximation revised for solids functional (PBEsol) \cite{PBEsol:2008} give similar results.

%nonetheless, we find critical inconsistencies between the ARPES and DFT calculations based on a correctly simulated  
%orthorhombic $Bmab$ phase of \LNO, which refute the main claim that ``\LNO\ has no pseudogap in the $d_{x^2-y^2}$ band, while it has an extra band of principally $d_{3z^2-r^2}$ orbital character, which presents a low temperature energy gap.''
%%
%We show that orthorhombic \LNO\ exhibits no pseudogap and that reconciliation between the ARPES data reported by 
%Li {\it et al.}\cite{Haoxiang:2017} and the DFT band structure requires that \LNO\ is monoclinic at low-temperature, indicating it 
%likely undergoes a displacive transition. 
%%\cite{priv_comm} The disappearance of the energy gap between 120 and 150?K is consistent with the resistivity curve in Fig. 4f, which display the anomaly at ~140?K. This indicates the likely connection of the energy gap opening to the phase transition found in the resistivity curve.
%Only the monoclinic phase exhibits the observed pseudogap.
%%which reconciles DFT calculations and ARPES data.

%\emph{Prediction from DFT calculations .} --- 
%\vspace{0.5em}
%\noindent
%\emph{{\bf Results}}\\
%\section{Results}
%\emph{{\bf Structural stability of \LNO.}}
%\subsection{Structural stability of \LNO}
\emph{Structural Stability.}---%of \emph{\LNO}.}---%
A survey of the available literature reveals that the crystal structure of \LNO\ is ambiguous.  While some authors report that the refined  structure of \LNO\ at 300~K is orthorhombic with space group $Bmab$\cite{Tkalich:1991, ling:2000}, others identified a monoclinic structure with space group $P2_1/a$ (hereafter referred to as monoclinic-I).\cite{Nagell:2015} In both interpretations, the crystal structure of \LNO\ is described using 4 formula units in the B-centered and primitive unit cells, respectively.
In  Supplementary Table~1 of Ref.\ \onlinecite{Supplemental_Note}, %\autoref{tab:tab1} 
we report the DFT-LDA relaxed atomic positions within the $Bmab$  space group using the experimental lattice parameters.\cite{ling:2000} 
We find that the positions for the free atomic Wyckoff orbits are in close agreement with the experimental values 
reported in Ref.\ \onlinecite{ling:2000}, confirming that the LDA functional is sufficiently accurate to describe the layered metallic nickelate \LNO. 

\begin{figure}[t]
\centering
\includegraphics[width=0.99\columnwidth]{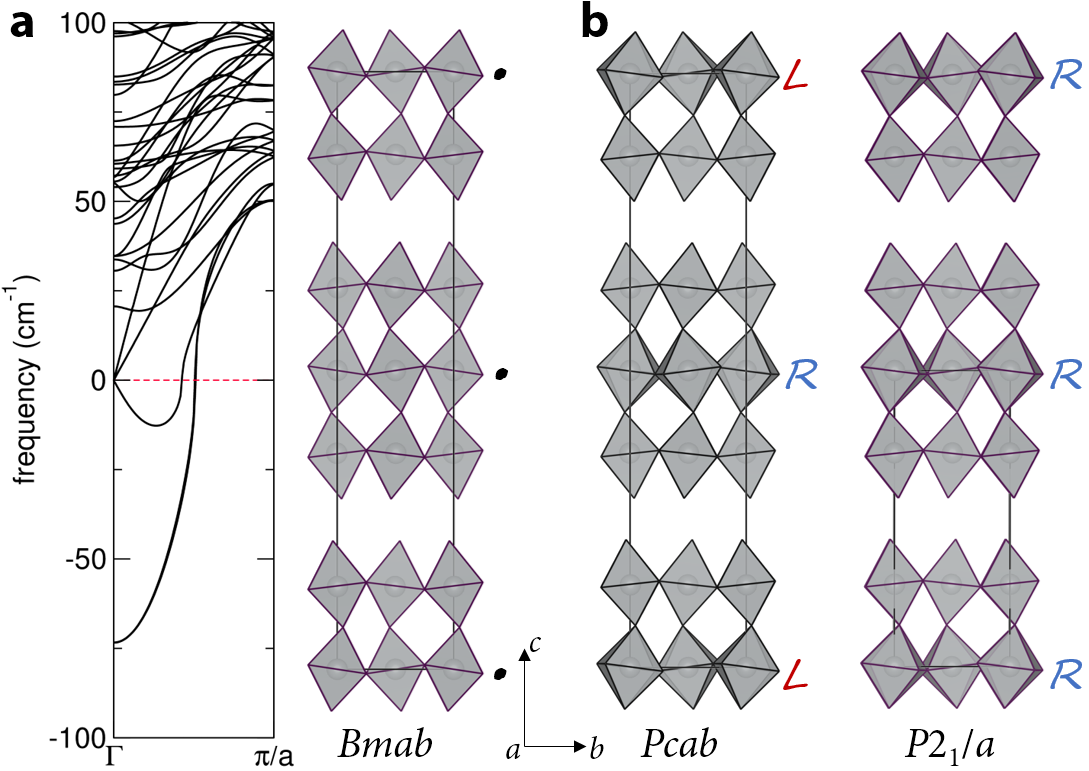}%\vspace{-6pt}
\caption{Structural polymorphs of \LNO.  (a) Phonon dispersions of the reported {\it Bmab} phase with its layered corner-connected NiO$_6$ crystal structure (La and oxygen atoms omitted for clarity). \cite{Tkalich:1991, ling:2000} (b) Crystal structures of the low temperature phases ($Pcab$ and monoclinic-II $P2_1/a$). The script letter indicate the sense of octahedral rotations about $c$ within the perovskite-layers; it is this change in stacking sequence of the central NiO$_2$ plane that results in the two competing polymorphs. The dot in panel (a) indicates no rotation. Note that the primitive $P2_1/a$ monoclinic-II phase is shown in a non-primitive cell to facilitate comparison.}
\label{fig1}
\end{figure}

To check the dynamical stability of the  $Bmab$ phase,  we performed first-principles phonon calculations. 
The result, shown in \autoref{fig1}a, indicates that the $Bmab$ phase is unstable at low temperatures.
It can transform to either the monoclinic $P2_1/a$ (monoclinic-II) polymorph (see Supplementary Table~2 in Ref.\ \onlinecite{Supplemental_Note} for structural information) %(\autoref{tab:tab2}) 
or the orthorhombic $Pcab$ polymorph (Supplementary Table~3) %(\autoref{tab:tab3}) 
with a primitive lattice (\autoref{fig1}b). 
Although the energy difference between these two structures is only %$\Delta E = 1.5$~meV 
1.5\,meV per formula unit (f.u.) in favor of the primitive orthorhombic structure, which makes them essentially degenerate at the LDA level, both phases are approximately  27\,meV/f.u.\ lower in energy than  
the  centered orthorhombic $Bmab$ crystal structure.
Note that our energy comparison is done by removing all symmetries from our calculations and using the same supercell volume and $k$-point sampling for the $Bmab$, $Pbca$, and $P2_1/a$ monoclinic-II phases.
 % is higher in energy of 27~meV per formula unit. 
Also, note that no group-subgroup relationship exists between the two predicted low-temperature phases and therefore 
the phase preference is likely to be strongly dependent on the  experimental conditions.

Next we evaluate the geometric-induced displacements of the $Pcab$ structure with respect to the $Bmab$ phase using a group-theoretical analysis.\cite{Orobengoa:ks5225, Perez-Mato:sh5107} 
These displacements reduce the $Pcab$ structure into a set of symmetry-adapted modes transforming as different irreducible representations (irreps) of the $Bmab$ phase. The loss of  symmetry derives from the appearance of an in-plane \emph{out-of-phase} rotation in the inner NiO$_2$ plane (\autoref{fig1}b) combined with small antipolar displacements along the $a$-axis of all the others atoms (not shown). The related distortion vector corresponds to irrep Y$_2^+$, which describes a phonon with $k = (1,0,0)$ wavevector and frequency $\omega=73.4i$~cm$^{-1}$ relative to the $Bmab$ phase (\autoref{fig1}a).
Similarly, we perform the group theoretical analysis on the $P2_1/a$ monoclinic-II phase. We find that the distortion vector corresponds to the irrep $\Gamma_2^+$ described by a phonon at $\Gamma$ and frequency $\omega=73.4i$~cm$^{-1}$ (\autoref{fig1}a). Differently from irrep Y$_2^+$, the zone-center instability exhibits an in-plane \emph{in-phase} rotation of the  NiO$_2$ plane between perovskite layers (\autoref{fig1}b) that results in the reduction of the periodicity of the crystal structure along the $c$ axis.

\begin{figure}[t]
\centering
\includegraphics[width=0.95\columnwidth]{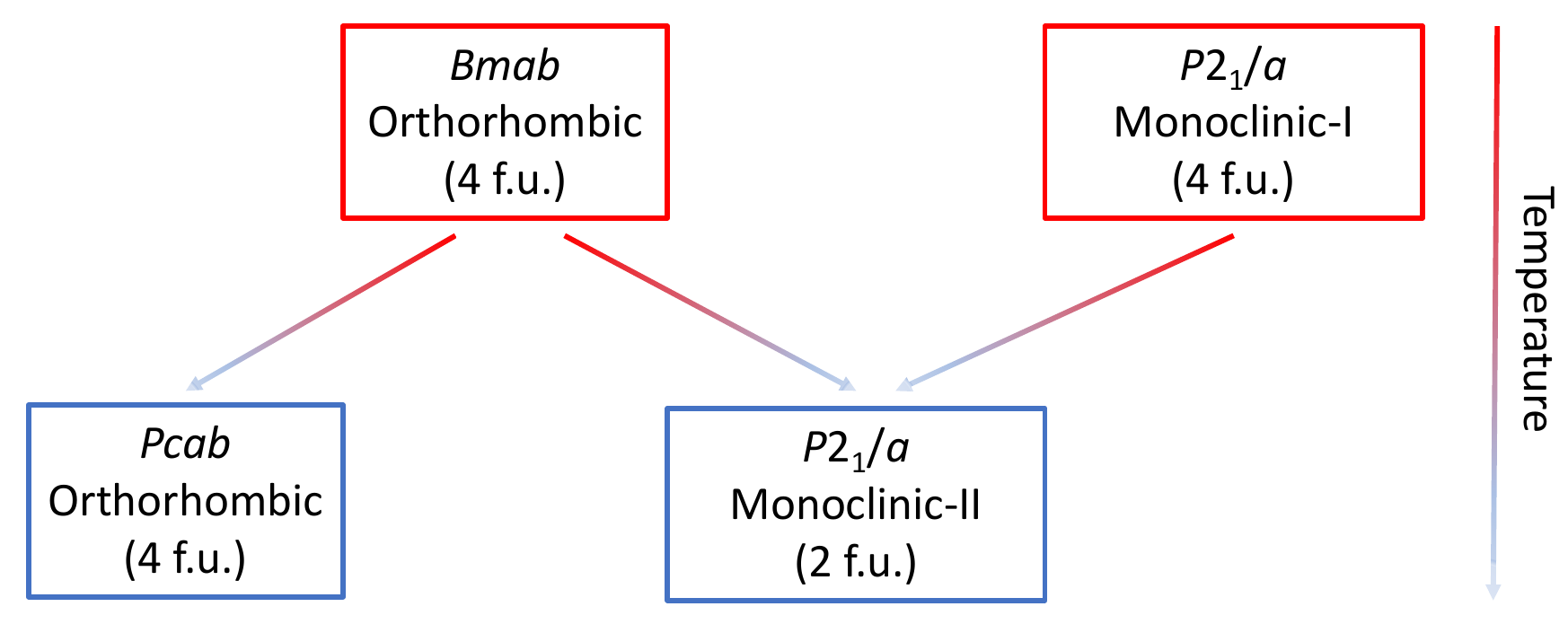}%\vspace{-7pt}
\caption{Symmetry reductions across the structural phase transition in \LNO. Although at 300\,K the orthorhombic $Bmab$ phase and monoclinic-I phase may coexist (upper row), at low temperature these phases to transform into the orthorhombic $Pbca$ and monoclinic-II phases (lower row).}
\label{fig2}
\end{figure}

Upon relaxation of the internal coordinates of the \mbox{monoclinic-I} $P2_1/a$ phase suggested by  Nagell {\it et al.}\cite{Nagell:2015}, we found that the structure transforms to the \mbox{monoclinic-II} $P2_1/a$ structure discussed above. Importantly, although the monoclinic-I phase is described using 4 formula units, the monoclinic-II $P2_1/a$ phase can be described using only 2 formula units. Therefore, being isomorphic, there is no allowed group-subgroup relationship between the two monoclinic phases and they can only be bridged via an intermediate phase.
Based on these findings, we propose that although at room temperature the $Bmab$ and monoclinic-I phases may coexist in \LNO,  the $Pbca$ and monoclinic-II phases are more likely to coexist at low temperatures (\autoref{fig2}). These subtle crystallographic distinctions will be important in understanding the temperature-dependent electronic  structure changes.

\begin{figure}[t]
\centering
\includegraphics[width=0.99\columnwidth]{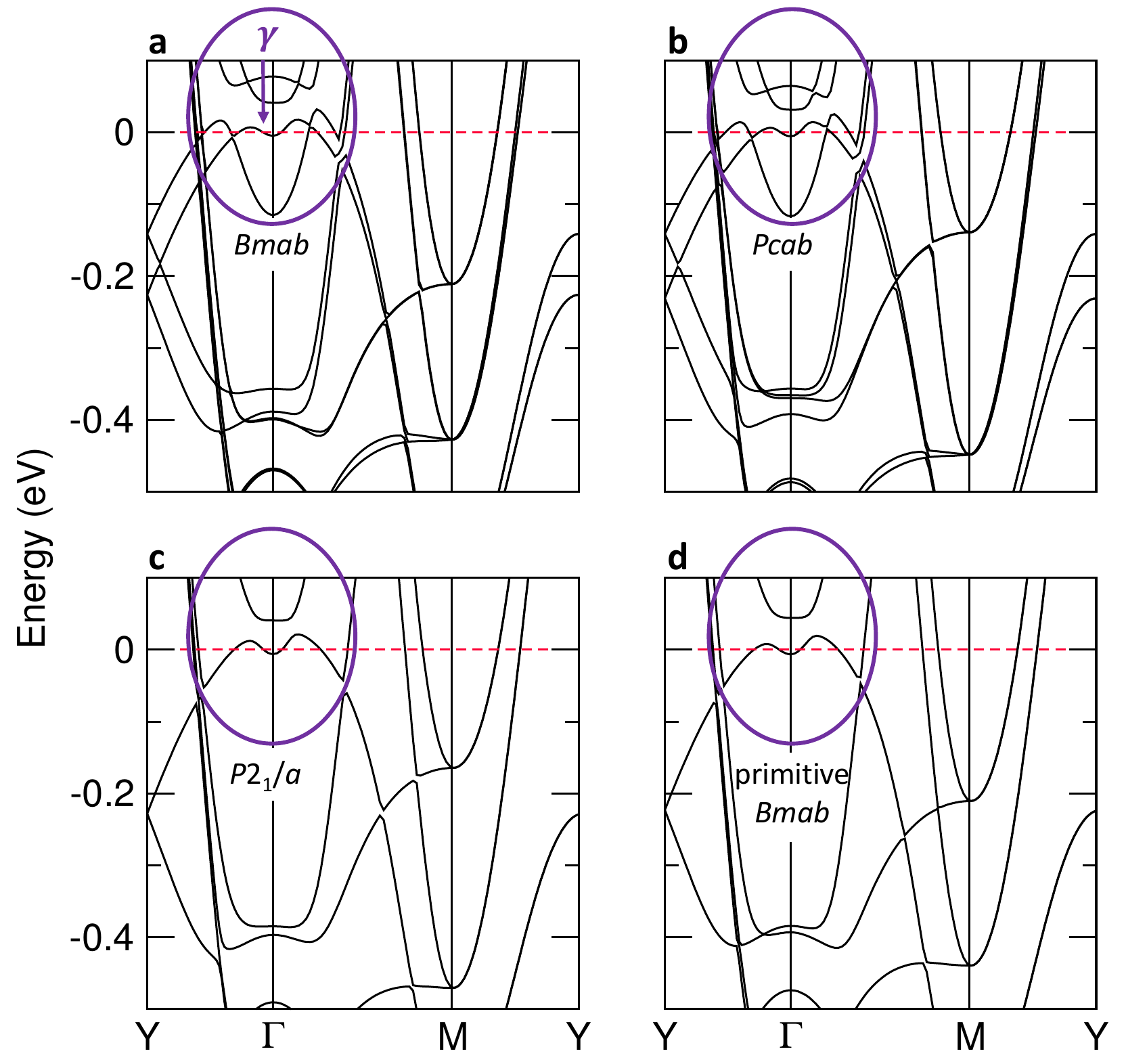}%\vspace{-7pt}
\caption{Electronic band structure of \LNO. Electronic properties of \LNO\ for the (a) orthorhombic high-temperature phase $Bmab$, (b) orthorhombic low-temperature phase $Pcab$, (c) monoclinic-II low temperature phase $P2_1/a$,   and (d) primitive $Bmab$ polymorph. The major low-energy differences are indicated by the open ring. Additional differences are clearly discernible at -0.4\,eV at the $\Gamma$ point.}
\label{fig3}
\end{figure}

%\vspace{0.5em}
%\noindent
%\emph{{\bf Electronic band structure of \LNO.}}
%\subsection{Electronic band structure of \LNO}
\emph{Electronic Structure.}---%
The electronic band structure for the orthorhombic high-temperature phase ($Bmab$) of \LNO\ is reported in \autoref{fig3}a. As expected, the electronic structure exhibits  band splitting  that arises from the multilayer coupling active in the trilayer perovskite. The splitting of the $\gamma$ band results in a hole pocket  centered at $\Gamma$ (\autoref{fig3}a), which  displays a linear band dispersion across the Fermi level and is consequently is ungapped. This feature is present in the 180~K APRES spectrum reported in Ref.~\onlinecite{Haoxiang:2017}. 
Note that the electronic band structure of the monoclinic-I  phase reported by Nagell {\it et al.}\cite{Nagell:2015} exhibits a similar feature (Supplementary Figure~1 of Ref.\ \onlinecite{Supplemental_Note}).
%Our electronic structure also qualitatively differs from that of  Ref.~\onlinecite{Haoxiang:2017}. \autoref{fig3}a indicates a number of bands are missing from Figure 2g of Ref.~\onlinecite{Haoxiang:2017}. For example, we find two bands originating at the Y point of the Brillouin zone at $\sim$200~meV below the Fermi level, whereas only one band is reported in Ref.~\onlinecite{Haoxiang:2017}.
%
Interesting, at low temperature these features are gapped. Li {\it et al.}\cite{Haoxiang:2017} suggest that this behavior can be connected  to the appearance of a potential charge-density waves (CDW) below the metal-to-metal transition around 140~K.\cite{seo:1996,Carvalho:2000}  However, as we describe next, our calculations suggest that this behavior is connected to the structural transition $Bmab\rightarrow P2_1/a$ and a property of the monoclinic-II phase of \LNO, which may or may not be CDW driven. This complex interplay between lattice instabilities and electronic structure in \LNO\ appear to be similar to those appearing in the metal-insulator transition of $t_{2g}$-electron systems.\cite{PhysRevB.91.195115}

The electronic structures for the orthorhombic ($Pcab$) and monoclinic-II ($P2_1/a$) low temperature phases are shown in  \autoref{fig3}b and \autoref{fig3}c, respectively.
Although the band structure for the primitive orthorhombic ($Pcab$) phase is similar to that of the centered orthorhombic high-temperature phase ($Bmab$), the electronic structure of the monoclinic-II $P2_1/a$ phase is markedly different. It does not exhibit band splitting due to the different periodicity of the crystal structure along the $c$ axis. Because the absence of band splitting, the dispersion of the $\gamma$ band with $d_{{3z^2-r^2}}$ orbital character turns flat and is in much better agreement with the pseudo-gap symmetrized spectrum taken at 24~K reported in Ref.~\onlinecite{Haoxiang:2017}.
Also, the lack of band splitting in the monoclinic $P2_1/a$ phase results in better agreement between the DFT band structure and the observed low temperature Fermi surface.\cite{Haoxiang:2017}

\begin{figure}[t]
\centering
\includegraphics[width=0.95\columnwidth]{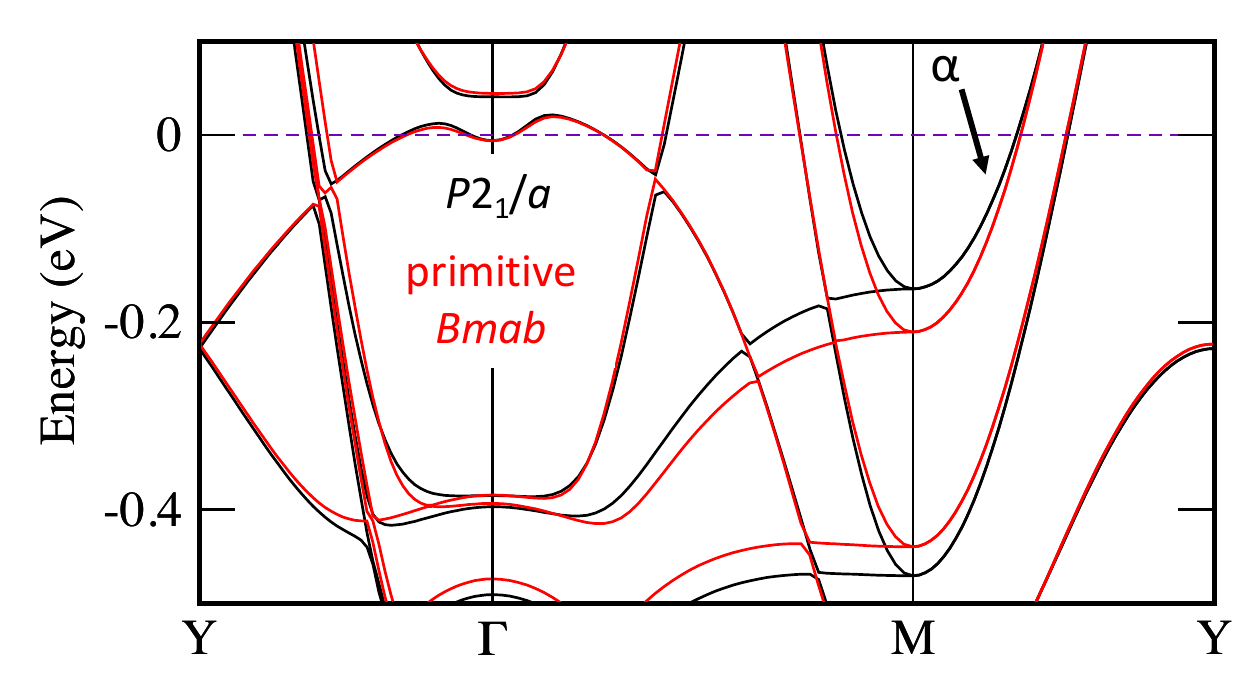}\vspace{-7pt}
\caption{Direct comparison of the electronic band structure of the monoclinic-II low temperature phase $P2_1/a$ (dark line) and primitive $Bmab$ polymorph (red line). Although the two band structures are similar, clear differences exist. In particular, the $\alpha$-band exhibits different dispersion for the two structure.}
\label{fig4}
\end{figure}

In \autoref{fig3}d  we show the electronic band structure of the primitive cell of the $Bmab$ structure.
The primitive cell has 2 formula units, see Supplementary Table~4 in Ref.\ \onlinecite{Supplemental_Note}, and was used by Li {\it et al.} for the band structure calculation in making comparisons with the ARPES data.\cite{Haoxiang:2017,priv_comm} 
Interesting, the low-energy electronic band structure of the monoclinic-II $P2_1/a$ (\autoref{fig3}c) is strikingly similar to that of the primitive cell of the $Bmab$ structure (\autoref{fig3}d).
After comparing the electronic band structures of the two phases, we find that the $\alpha$-band close to the M point for the primitive $Cmca$ is more dispersive than the $\alpha$-band of the monoclinic-II $P2_1/a$ phase (\autoref{fig4}). This results in a different value of the mass enhancement for the monoclinic-II $P2_1/a$ that we quantify to be around 1.7, which is  23\% less than the value calculated in Ref.~\onlinecite{Haoxiang:2017}. 
Additional distinctions between the two phases can be found at lower energy (near -0.4\,eV) beginning at $\Gamma$ and evolving into the BZ. 
%

%\section{Conclusion}
\emph{Conclusions.}---%
In summary, we propose that the reported high-temperature phases of \LNO, which both have 4 formula units, are unstable at low temperature and transform to a monoclinic $P2_1/a$ phase that can be described using 2 formula unit. 
Moreover, the electronic structure of the new monoclinic-II $P2_1/a$ phase is in better agreement with available experimental ARPES data.\cite{Haoxiang:2017} 
An important point to emphasize is that the interaxial angles and lattice parameters of the primitive cell of the $Bmab$ phase are nearly identical to those of the monoclinic-II structure.\cite{Supplemental_Note}
This is purely accidental as the low-symmetry phase obtained from the phonons could exhibit one of 13 monoclinic symmetries. 

Our results highlight the need for careful consideration of crystal-structure details when studying 
temperature dependent electronic structure changes.
Structural assignments of trilayered perovskites based only on crystal system can lead to considerable misinterpretation and the loss of key physics, especially when changes in electronic structure are less stark, e.g., at a weak metal-to-metal transition. 
Last, this  finding suggests the resistivity anomaly detected at $\sim$140~K  in Ref.~\onlinecite{Haoxiang:2017} could have a nontrivial structural component associated with it.

%\section*{Acknowledgement}
%
\emph{Acknowledgement.}---%
D.P.\ and J.M.R.\ acknowledge the ARO (W911NF-15-1-0017) and NSF (DMR-1729303) for financial support and the DOD-HPCMP for computational resources.

\bibliography{puggioni}

\begin{thebibliography}{28}%
\makeatletter
\providecommand \@ifxundefined [1]{%
 \@ifx{#1\undefined}
}%
\providecommand \@ifnum [1]{%
 \ifnum #1\expandafter \@firstoftwo
 \else \expandafter \@secondoftwo
 \fi
}%
\providecommand \@ifx [1]{%
 \ifx #1\expandafter \@firstoftwo
 \else \expandafter \@secondoftwo
 \fi
}%
\providecommand \natexlab [1]{#1}%
\providecommand \enquote  [1]{``#1''}%
\providecommand \bibnamefont  [1]{#1}%
\providecommand \bibfnamefont [1]{#1}%
\providecommand \citenamefont [1]{#1}%
\providecommand \href@noop [0]{\@secondoftwo}%
\providecommand \href [0]{\begingroup \@sanitize@url \@href}%
\providecommand \@href[1]{\@@startlink{#1}\@@href}%
\providecommand \@@href[1]{\endgroup#1\@@endlink}%
\providecommand \@sanitize@url [0]{\catcode `\\12\catcode `\$12\catcode
  `\&12\catcode `\#12\catcode `\^12\catcode `\_12\catcode `\%12\relax}%
\providecommand \@@startlink[1]{}%
\providecommand \@@endlink[0]{}%
\providecommand \url  [0]{\begingroup\@sanitize@url \@url }%
\providecommand \@url [1]{\endgroup\@href {#1}{\urlprefix }}%
\providecommand \urlprefix  [0]{URL }%
\providecommand \Eprint [0]{\href }%
\providecommand \doibase [0]{http://dx.doi.org/}%
\providecommand \selectlanguage [0]{\@gobble}%
\providecommand \bibinfo  [0]{\@secondoftwo}%
\providecommand \bibfield  [0]{\@secondoftwo}%
\providecommand \translation [1]{[#1]}%
\providecommand \BibitemOpen [0]{}%
\providecommand \bibitemStop [0]{}%
\providecommand \bibitemNoStop [0]{.\EOS\space}%
\providecommand \EOS [0]{\spacefactor3000\relax}%
\providecommand \BibitemShut  [1]{\csname bibitem#1\endcsname}%
\let\auto@bib@innerbib\@empty
%</preamble>
\bibitem [{\citenamefont {Bednorz}\ and\ \citenamefont
  {M{\"u}ller}(1986)}]{Bednorz/Muller:1986}%
  \BibitemOpen
  \bibfield  {author} {\bibinfo {author} {\bibfnamefont {J.~G.}\ \bibnamefont
  {Bednorz}}\ and\ \bibinfo {author} {\bibfnamefont {K.~A.}\ \bibnamefont
  {M{\"u}ller}},\ }\href@noop {} {\bibfield  {journal} {\bibinfo  {journal}
  {Zeitschrift f{\"u}r Physik B Condensed Matter}\ }\textbf {\bibinfo {volume}
  {64}},\ \bibinfo {pages} {189} (\bibinfo {year} {1986})}\BibitemShut
  {NoStop}%
\bibitem [{hea(2017)}]{heatingup:2017}%
  \BibitemOpen
  \href {https://journals.aps.org/prl/heating-up-of-superconductors} {\enquote
  {\bibinfo {title} {Heating up of superconductors},}\ } (\bibinfo {year}
  {2017}),\ \bibinfo {note} {this \emph{Phys.\ Rev.\ Lett.} collection marks
  the 30th anniversary of the discovery of high-temperature
  superconductors.}\BibitemShut {Stop}%
\bibitem [{\citenamefont {Maeno}\ \emph {et~al.}(1994)\citenamefont {Maeno},
  \citenamefont {Hashimoto}, \citenamefont {Yoshida}, \citenamefont
  {Nishizaki}, \citenamefont {Fujita}, \citenamefont {Bednorz},\ and\
  \citenamefont {Lichtenberg}}]{Maeno_et_al:1994}%
  \BibitemOpen
  \bibfield  {author} {\bibinfo {author} {\bibfnamefont {Y.}~\bibnamefont
  {Maeno}}, \bibinfo {author} {\bibfnamefont {H.}~\bibnamefont {Hashimoto}},
  \bibinfo {author} {\bibfnamefont {K.}~\bibnamefont {Yoshida}}, \bibinfo
  {author} {\bibfnamefont {S.}~\bibnamefont {Nishizaki}}, \bibinfo {author}
  {\bibfnamefont {T.}~\bibnamefont {Fujita}}, \bibinfo {author} {\bibfnamefont
  {J.~G.}\ \bibnamefont {Bednorz}}, \ and\ \bibinfo {author} {\bibfnamefont
  {F.}~\bibnamefont {Lichtenberg}},\ }\href@noop {} {\bibfield  {journal}
  {\bibinfo  {journal} {Nature}\ }\textbf {\bibinfo {volume} {372}},\ \bibinfo
  {pages} {532} (\bibinfo {year} {1994})}\BibitemShut {NoStop}%
\bibitem [{\citenamefont {Takada}\ \emph {et~al.}(2003)\citenamefont {Takada},
  \citenamefont {Sakurai}, \citenamefont {Takayama-Muromachi}, \citenamefont
  {Izumi}, \citenamefont {Dilanian},\ and\ \citenamefont
  {Sasaki}}]{takada:2003}%
  \BibitemOpen
  \bibfield  {author} {\bibinfo {author} {\bibfnamefont {K.}~\bibnamefont
  {Takada}}, \bibinfo {author} {\bibfnamefont {H.}~\bibnamefont {Sakurai}},
  \bibinfo {author} {\bibfnamefont {E.}~\bibnamefont {Takayama-Muromachi}},
  \bibinfo {author} {\bibfnamefont {F.}~\bibnamefont {Izumi}}, \bibinfo
  {author} {\bibfnamefont {R.~A.}\ \bibnamefont {Dilanian}}, \ and\ \bibinfo
  {author} {\bibfnamefont {T.}~\bibnamefont {Sasaki}},\ }\href@noop {}
  {\bibfield  {journal} {\bibinfo  {journal} {Nature}\ }\textbf {\bibinfo
  {volume} {422}},\ \bibinfo {pages} {53 EP } (\bibinfo {year}
  {2003})}\BibitemShut {NoStop}%
\bibitem [{\citenamefont {Anisimov}\ \emph {et~al.}(1999)\citenamefont
  {Anisimov}, \citenamefont {Bukhvalov},\ and\ \citenamefont
  {Rice}}]{Anisimov/Bukhvalov/Rice:1999}%
  \BibitemOpen
  \bibfield  {author} {\bibinfo {author} {\bibfnamefont {V.~I.}\ \bibnamefont
  {Anisimov}}, \bibinfo {author} {\bibfnamefont {D.}~\bibnamefont {Bukhvalov}},
  \ and\ \bibinfo {author} {\bibfnamefont {T.~M.}\ \bibnamefont {Rice}},\
  }\href@noop {} {\bibfield  {journal} {\bibinfo  {journal} {Physical Review
  B}\ }\textbf {\bibinfo {volume} {59}},\ \bibinfo {pages} {7901} (\bibinfo
  {year} {1999})}\BibitemShut {NoStop}%
\bibitem [{\citenamefont {Botana}\ \emph {et~al.}(2017)\citenamefont {Botana},
  \citenamefont {Pardo},\ and\ \citenamefont
  {Norman}}]{PhysRevMaterials.1.021801}%
  \BibitemOpen
  \bibfield  {author} {\bibinfo {author} {\bibfnamefont {A.~S.}\ \bibnamefont
  {Botana}}, \bibinfo {author} {\bibfnamefont {V.}~\bibnamefont {Pardo}}, \
  and\ \bibinfo {author} {\bibfnamefont {M.~R.}\ \bibnamefont {Norman}},\
  }\href {\doibase 10.1103/PhysRevMaterials.1.021801} {\bibfield  {journal}
  {\bibinfo  {journal} {Phys. Rev. Materials}\ }\textbf {\bibinfo {volume}
  {1}},\ \bibinfo {pages} {021801} (\bibinfo {year} {2017})}\BibitemShut
  {NoStop}%
\bibitem [{\citenamefont {Zhong}\ \emph {et~al.}(2017)\citenamefont {Zhong},
  \citenamefont {Winn}, \citenamefont {Gu}, \citenamefont {Reznik},\ and\
  \citenamefont {Tranquada}}]{PhysRevLett.118.177601}%
  \BibitemOpen
  \bibfield  {author} {\bibinfo {author} {\bibfnamefont {R.}~\bibnamefont
  {Zhong}}, \bibinfo {author} {\bibfnamefont {B.~L.}\ \bibnamefont {Winn}},
  \bibinfo {author} {\bibfnamefont {G.}~\bibnamefont {Gu}}, \bibinfo {author}
  {\bibfnamefont {D.}~\bibnamefont {Reznik}}, \ and\ \bibinfo {author}
  {\bibfnamefont {J.~M.}\ \bibnamefont {Tranquada}},\ }\href {\doibase
  10.1103/PhysRevLett.118.177601} {\bibfield  {journal} {\bibinfo  {journal}
  {Phys. Rev. Lett.}\ }\textbf {\bibinfo {volume} {118}},\ \bibinfo {pages}
  {177601} (\bibinfo {year} {2017})}\BibitemShut {NoStop}%
\bibitem [{\citenamefont {Kajimoto}\ \emph {et~al.}(2003)\citenamefont
  {Kajimoto}, \citenamefont {Ishizaka}, \citenamefont {Yoshizawa},\ and\
  \citenamefont {Tokura}}]{PhysRevB.67.014511}%
  \BibitemOpen
  \bibfield  {author} {\bibinfo {author} {\bibfnamefont {R.}~\bibnamefont
  {Kajimoto}}, \bibinfo {author} {\bibfnamefont {K.}~\bibnamefont {Ishizaka}},
  \bibinfo {author} {\bibfnamefont {H.}~\bibnamefont {Yoshizawa}}, \ and\
  \bibinfo {author} {\bibfnamefont {Y.}~\bibnamefont {Tokura}},\ }\href
  {\doibase 10.1103/PhysRevB.67.014511} {\bibfield  {journal} {\bibinfo
  {journal} {Phys. Rev. B}\ }\textbf {\bibinfo {volume} {67}},\ \bibinfo
  {pages} {014511} (\bibinfo {year} {2003})}\BibitemShut {NoStop}%
\bibitem [{\citenamefont {Zhang}\ and\ \citenamefont
  {Greenblatt}(1995)}]{ZHANG1995236}%
  \BibitemOpen
  \bibfield  {author} {\bibinfo {author} {\bibfnamefont {Z.}~\bibnamefont
  {Zhang}}\ and\ \bibinfo {author} {\bibfnamefont {M.}~\bibnamefont
  {Greenblatt}},\ }\href {\doibase https://doi.org/10.1006/jssc.1995.1269}
  {\bibfield  {journal} {\bibinfo  {journal} {Journal of Solid State
  Chemistry}\ }\textbf {\bibinfo {volume} {117}},\ \bibinfo {pages} {236 }
  (\bibinfo {year} {1995})}\BibitemShut {NoStop}%
\bibitem [{\citenamefont {Tkalich}\ \emph {et~al.}(1991)\citenamefont
  {Tkalich}, \citenamefont {Glazkov}, \citenamefont {Somenkov}, \citenamefont
  {Shilshtein}, \citenamefont {Karkin},\ and\ \citenamefont
  {Mirmelshtein}}]{Tkalich:1991}%
  \BibitemOpen
  \bibfield  {author} {\bibinfo {author} {\bibfnamefont {A.~K.}\ \bibnamefont
  {Tkalich}}, \bibinfo {author} {\bibfnamefont {V.~P.}\ \bibnamefont
  {Glazkov}}, \bibinfo {author} {\bibfnamefont {V.~A.}\ \bibnamefont
  {Somenkov}}, \bibinfo {author} {\bibfnamefont {S.~S.}\ \bibnamefont
  {Shilshtein}}, \bibinfo {author} {\bibfnamefont {A.~E.}\ \bibnamefont
  {Karkin}}, \ and\ \bibinfo {author} {\bibfnamefont {A.~V.}\ \bibnamefont
  {Mirmelshtein}},\ }\href@noop {} {\bibfield  {journal} {\bibinfo  {journal}
  {{Superconductivity}}\ }\textbf {\bibinfo {volume} {4}},\ \bibinfo {pages}
  {2280} (\bibinfo {year} {1991})}\BibitemShut {NoStop}%
\bibitem [{\citenamefont {{H.\ Li \emph{et al}.}}(2017)}]{Haoxiang:2017}%
  \BibitemOpen
  \bibfield  {author} {\bibinfo {author} {\bibnamefont {{H.\ Li \emph{et
  al}.}}},\ }\href@noop {} {\bibfield  {journal} {\bibinfo  {journal} {Nat.
  Commun.}\ }\textbf {\bibinfo {volume} {8}},\ \bibinfo {pages} {704} (\bibinfo
  {year} {2017})}\BibitemShut {NoStop}%
\bibitem [{\citenamefont {Rabe}\ \emph {et~al.}(1992)\citenamefont {Rabe},
  \citenamefont {Phillips}, \citenamefont {Villars},\ and\ \citenamefont
  {Brown}}]{PhysRevB.45.7650}%
  \BibitemOpen
  \bibfield  {author} {\bibinfo {author} {\bibfnamefont {K.~M.}\ \bibnamefont
  {Rabe}}, \bibinfo {author} {\bibfnamefont {J.~C.}\ \bibnamefont {Phillips}},
  \bibinfo {author} {\bibfnamefont {P.}~\bibnamefont {Villars}}, \ and\
  \bibinfo {author} {\bibfnamefont {I.~D.}\ \bibnamefont {Brown}},\ }\href
  {\doibase 10.1103/PhysRevB.45.7650} {\bibfield  {journal} {\bibinfo
  {journal} {Phys. Rev. B}\ }\textbf {\bibinfo {volume} {45}},\ \bibinfo
  {pages} {7650} (\bibinfo {year} {1992})}\BibitemShut {NoStop}%
\bibitem [{\citenamefont {Leonov}\ \emph {et~al.}(2015)\citenamefont {Leonov},
  \citenamefont {Anisimov},\ and\ \citenamefont
  {Vollhardt}}]{PhysRevB.91.195115}%
  \BibitemOpen
  \bibfield  {author} {\bibinfo {author} {\bibfnamefont {I.}~\bibnamefont
  {Leonov}}, \bibinfo {author} {\bibfnamefont {V.~I.}\ \bibnamefont
  {Anisimov}}, \ and\ \bibinfo {author} {\bibfnamefont {D.}~\bibnamefont
  {Vollhardt}},\ }\href {\doibase 10.1103/PhysRevB.91.195115} {\bibfield
  {journal} {\bibinfo  {journal} {Phys. Rev. B}\ }\textbf {\bibinfo {volume}
  {91}},\ \bibinfo {pages} {195115} (\bibinfo {year} {2015})}\BibitemShut
  {NoStop}%
\bibitem [{\citenamefont {Motizuki}(2012)}]{motizuki2012structural}%
  \BibitemOpen
  \bibfield  {author} {\bibinfo {author} {\bibfnamefont {K.}~\bibnamefont
  {Motizuki}},\ }\href {https://books.google.com/books?id=G-0ZCQAAQBAJ} {\emph
  {\bibinfo {title} {Structural Phase Transitions in Layered Transition Metal
  Compounds}}},\ Physics and Chemistry of Materials with A\ (\bibinfo
  {publisher} {Springer Netherlands},\ \bibinfo {year} {2012})\BibitemShut
  {NoStop}%
\bibitem [{\citenamefont {\emph{et al}.}()}]{priv_comm}%
  \BibitemOpen
  \bibfield  {author} {\bibinfo {author} {\bibfnamefont {H.~L.}\ \bibnamefont
  {\emph{et al}.}},\ }\href@noop {} {}\bibinfo {howpublished} {personal
  communication}\BibitemShut {NoStop}%
\bibitem [{\citenamefont {Perdew}\ and\ \citenamefont
  {Zunger}(1981)}]{Perdew/Zunger:1981}%
  \BibitemOpen
  \bibfield  {author} {\bibinfo {author} {\bibfnamefont {J.~P.}\ \bibnamefont
  {Perdew}}\ and\ \bibinfo {author} {\bibfnamefont {A.}~\bibnamefont
  {Zunger}},\ }\href@noop {} {\bibfield  {journal} {\bibinfo  {journal}
  {Physical Review B}\ }\textbf {\bibinfo {volume} {23}},\ \bibinfo {pages}
  {5048} (\bibinfo {year} {1981})}\BibitemShut {NoStop}%
\bibitem [{\citenamefont {Kresse}\ and\ \citenamefont
  {Furthm\"uller}(1996)}]{Kresse/Furthmuller:1996b}%
  \BibitemOpen
  \bibfield  {author} {\bibinfo {author} {\bibfnamefont {G.}~\bibnamefont
  {Kresse}}\ and\ \bibinfo {author} {\bibfnamefont {J.}~\bibnamefont
  {Furthm\"uller}},\ }\href@noop {} {\bibfield  {journal} {\bibinfo  {journal}
  {Computational Materials Science}\ }\textbf {\bibinfo {volume} {6}},\
  \bibinfo {pages} {15 } (\bibinfo {year} {1996})}\BibitemShut {NoStop}%
\bibitem [{\citenamefont {Bl\"ochl}\ \emph {et~al.}(1994)\citenamefont
  {Bl\"ochl}, \citenamefont {Jepsen},\ and\ \citenamefont
  {Andersen}}]{Blochl/Jepsen/Andersen:1994}%
  \BibitemOpen
  \bibfield  {author} {\bibinfo {author} {\bibfnamefont {P.~E.}\ \bibnamefont
  {Bl\"ochl}}, \bibinfo {author} {\bibfnamefont {O.}~\bibnamefont {Jepsen}}, \
  and\ \bibinfo {author} {\bibfnamefont {O.~K.}\ \bibnamefont {Andersen}},\
  }\href@noop {} {\bibfield  {journal} {\bibinfo  {journal} {Physical Review
  B}\ }\textbf {\bibinfo {volume} {49}},\ \bibinfo {pages} {16223} (\bibinfo
  {year} {1994})}\BibitemShut {NoStop}%
\bibitem [{\citenamefont {Monkhorst}\ and\ \citenamefont
  {Pack}(1976)}]{Monkhorst/Pack:1976}%
  \BibitemOpen
  \bibfield  {author} {\bibinfo {author} {\bibfnamefont {H.~J.}\ \bibnamefont
  {Monkhorst}}\ and\ \bibinfo {author} {\bibfnamefont {J.~D.}\ \bibnamefont
  {Pack}},\ }\href@noop {} {\bibfield  {journal} {\bibinfo  {journal} {Physical
  Review B}\ }\textbf {\bibinfo {volume} {13}},\ \bibinfo {pages} {5188}
  (\bibinfo {year} {1976})}\BibitemShut {NoStop}%
\bibitem [{\citenamefont {Orobengoa}\ \emph {et~al.}(2009)\citenamefont
  {Orobengoa}, \citenamefont {Capillas}, \citenamefont {Aroyo},\ and\
  \citenamefont {Perez-Mato}}]{Orobengoa:ks5225}%
  \BibitemOpen
  \bibfield  {author} {\bibinfo {author} {\bibfnamefont {D.}~\bibnamefont
  {Orobengoa}}, \bibinfo {author} {\bibfnamefont {C.}~\bibnamefont {Capillas}},
  \bibinfo {author} {\bibfnamefont {M.~I.}\ \bibnamefont {Aroyo}}, \ and\
  \bibinfo {author} {\bibfnamefont {J.~M.}\ \bibnamefont {Perez-Mato}},\
  }\href@noop {} {\bibfield  {journal} {\bibinfo  {journal} {Journal of Applied
  Crystallography}\ }\textbf {\bibinfo {volume} {42}},\ \bibinfo {pages} {820}
  (\bibinfo {year} {2009})}\BibitemShut {NoStop}%
\bibitem [{\citenamefont {Perez-Mato}\ \emph {et~al.}(2010)\citenamefont
  {Perez-Mato}, \citenamefont {Orobengoa},\ and\ \citenamefont
  {Aroyo}}]{Perez-Mato:sh5107}%
  \BibitemOpen
  \bibfield  {author} {\bibinfo {author} {\bibfnamefont {J.~M.}\ \bibnamefont
  {Perez-Mato}}, \bibinfo {author} {\bibfnamefont {D.}~\bibnamefont
  {Orobengoa}}, \ and\ \bibinfo {author} {\bibfnamefont {M.~I.}\ \bibnamefont
  {Aroyo}},\ }\href@noop {} {\bibfield  {journal} {\bibinfo  {journal} {Acta
  Crystallographica Section A}\ }\textbf {\bibinfo {volume} {66}},\ \bibinfo
  {pages} {558} (\bibinfo {year} {2010})}\BibitemShut {NoStop}%
\bibitem [{\citenamefont {Perdew}\ \emph {et~al.}(1996)\citenamefont {Perdew},
  \citenamefont {Burke},\ and\ \citenamefont
  {Ernzerhof}}]{Perdew/Burke/Ernzerhof:1996}%
  \BibitemOpen
  \bibfield  {author} {\bibinfo {author} {\bibfnamefont {J.~P.}\ \bibnamefont
  {Perdew}}, \bibinfo {author} {\bibfnamefont {K.}~\bibnamefont {Burke}}, \
  and\ \bibinfo {author} {\bibfnamefont {M.}~\bibnamefont {Ernzerhof}},\
  }\href@noop {} {\bibfield  {journal} {\bibinfo  {journal} {Physical Review
  Letters}\ }\textbf {\bibinfo {volume} {77}},\ \bibinfo {pages} {3865}
  (\bibinfo {year} {1996})}\BibitemShut {NoStop}%
\bibitem [{\citenamefont {Perdew}\ \emph {et~al.}(2008)\citenamefont {Perdew},
  \citenamefont {Ruzsinszky}, \citenamefont {Csonka}, \citenamefont {Vydrov},
  \citenamefont {Scuseria}, \citenamefont {Constantin}, \citenamefont {Zhou},\
  and\ \citenamefont {Burke}}]{PBEsol:2008}%
  \BibitemOpen
  \bibfield  {author} {\bibinfo {author} {\bibfnamefont {J.~P.}\ \bibnamefont
  {Perdew}}, \bibinfo {author} {\bibfnamefont {A.}~\bibnamefont {Ruzsinszky}},
  \bibinfo {author} {\bibfnamefont {G.~I.}\ \bibnamefont {Csonka}}, \bibinfo
  {author} {\bibfnamefont {O.~A.}\ \bibnamefont {Vydrov}}, \bibinfo {author}
  {\bibfnamefont {G.~E.}\ \bibnamefont {Scuseria}}, \bibinfo {author}
  {\bibfnamefont {L.~A.}\ \bibnamefont {Constantin}}, \bibinfo {author}
  {\bibfnamefont {X.}~\bibnamefont {Zhou}}, \ and\ \bibinfo {author}
  {\bibfnamefont {K.}~\bibnamefont {Burke}},\ }\href@noop {} {\bibfield
  {journal} {\bibinfo  {journal} {Physical Review Letters}\ }\textbf {\bibinfo
  {volume} {100}},\ \bibinfo {pages} {136406} (\bibinfo {year}
  {2008})}\BibitemShut {NoStop}%
\bibitem [{\citenamefont {Ling}\ \emph {et~al.}(2000)\citenamefont {Ling},
  \citenamefont {Argyriou}, \citenamefont {Wu},\ and\ \citenamefont
  {Neumeier}}]{ling:2000}%
  \BibitemOpen
  \bibfield  {author} {\bibinfo {author} {\bibfnamefont {C.~D.}\ \bibnamefont
  {Ling}}, \bibinfo {author} {\bibfnamefont {D.~N.}\ \bibnamefont {Argyriou}},
  \bibinfo {author} {\bibfnamefont {G.}~\bibnamefont {Wu}}, \ and\ \bibinfo
  {author} {\bibfnamefont {J.}~\bibnamefont {Neumeier}},\ }\href@noop {}
  {\bibfield  {journal} {\bibinfo  {journal} {Journal of Solid State
  Chemistry}\ }\textbf {\bibinfo {volume} {152}},\ \bibinfo {pages} {517 }
  (\bibinfo {year} {2000})}\BibitemShut {NoStop}%
\bibitem [{\citenamefont {Nagell}\ \emph {et~al.}(2015)\citenamefont {Nagell},
  \citenamefont {Kumar}, \citenamefont {S{\o}rby}, \citenamefont
  {Fjellv{\aa}g},\ and\ \citenamefont {Sj{\aa}stad}}]{Nagell:2015}%
  \BibitemOpen
  \bibfield  {author} {\bibinfo {author} {\bibfnamefont {M.~U.}\ \bibnamefont
  {Nagell}}, \bibinfo {author} {\bibfnamefont {S.}~\bibnamefont {Kumar}},
  \bibinfo {author} {\bibfnamefont {M.~H.}\ \bibnamefont {S{\o}rby}}, \bibinfo
  {author} {\bibfnamefont {H.}~\bibnamefont {Fjellv{\aa}g}}, \ and\ \bibinfo
  {author} {\bibfnamefont {A.~O.}\ \bibnamefont {Sj{\aa}stad}},\ }\href@noop {}
  {\bibfield  {journal} {\bibinfo  {journal} {Phase Transitions}\ }\textbf
  {\bibinfo {volume} {88}},\ \bibinfo {pages} {979} (\bibinfo {year}
  {2015})}\BibitemShut {NoStop}%
\bibitem [{Sup()}]{Supplemental_Note}%
  \BibitemOpen
  \href@noop {} {\ }\bibinfo {note} {See Supplemental Material at [URL will be
  inserted by publisher] for crystal structural information and additional
  electronic structures.}\BibitemShut {Stop}%
\bibitem [{\citenamefont {Seo}\ \emph {et~al.}(1996)\citenamefont {Seo},
  \citenamefont {Liang}, \citenamefont {Whangbo}, \citenamefont {Zhang},\ and\
  \citenamefont {Greenblatt}}]{seo:1996}%
  \BibitemOpen
  \bibfield  {author} {\bibinfo {author} {\bibfnamefont {D.-K.}\ \bibnamefont
  {Seo}}, \bibinfo {author} {\bibfnamefont {W.}~\bibnamefont {Liang}}, \bibinfo
  {author} {\bibfnamefont {M.-H.}\ \bibnamefont {Whangbo}}, \bibinfo {author}
  {\bibfnamefont {Z.}~\bibnamefont {Zhang}}, \ and\ \bibinfo {author}
  {\bibfnamefont {M.}~\bibnamefont {Greenblatt}},\ }\href@noop {} {\bibfield
  {journal} {\bibinfo  {journal} {Inorganic Chemistry}\ }\textbf {\bibinfo
  {volume} {35}},\ \bibinfo {pages} {6396} (\bibinfo {year}
  {1996})}\BibitemShut {NoStop}%
\bibitem [{\citenamefont {Carvalho}\ \emph {et~al.}(2000)\citenamefont
  {Carvalho}, \citenamefont {Cruz}, \citenamefont {Wattiaux}, \citenamefont
  {Bassat}, \citenamefont {Costa},\ and\ \citenamefont
  {Godinho}}]{Carvalho:2000}%
  \BibitemOpen
  \bibfield  {author} {\bibinfo {author} {\bibfnamefont {M.~D.}\ \bibnamefont
  {Carvalho}}, \bibinfo {author} {\bibfnamefont {M.~M.}\ \bibnamefont {Cruz}},
  \bibinfo {author} {\bibfnamefont {A.}~\bibnamefont {Wattiaux}}, \bibinfo
  {author} {\bibfnamefont {J.~M.}\ \bibnamefont {Bassat}}, \bibinfo {author}
  {\bibfnamefont {F.~M.~A.}\ \bibnamefont {Costa}}, \ and\ \bibinfo {author}
  {\bibfnamefont {M.}~\bibnamefont {Godinho}},\ }\href@noop {} {\bibfield
  {journal} {\bibinfo  {journal} {Journal of Applied Physics}\ }\textbf
  {\bibinfo {volume} {88}},\ \bibinfo {pages} {544} (\bibinfo {year}
  {2000})}\BibitemShut {NoStop}%
\end{thebibliography}%

%\newpage 
\clearpage

\onecolumngrid

\begin{center}
\large{\textsc{supporting information}}
\end{center}

%\vfill
\begingroup
%\squeezetable
\begin{table*}[h]
\begin{ruledtabular}
\centering
\caption{\label{tab:tab1}Crystallographic parameters for orthorhombic $Bmab$  \LNO\ from DFT-LDA. The lattice parameters are fixed to the experimental values reported in Ref. \onlinecite{ling:2000}.} %and
\begin{tabular}{lcccc}%
\multicolumn{5}{l}{$Bmab$ (64)} $a = 5.41327$~\AA,  $b = 5.46233$~\AA,  $c = 27.96049$~\AA \\
Atom    &       Wyck.\ Site     & $x$   & $y$   & $z$   \\
\hline
La(1)        	& $8f$	&   	0 		& 	-0.0018  	&		0.4331  \\
La(2)         &$8f$   	&	0  		&	-0.0116  	&		0.3022  \\
Ni(1)         &$4b$   	&	0 		&		0  	&		0  \\
Ni(2)        	&$8f$   	& 	0          	&	-0.0033 	& 		0.1384 \\
O(1)      	&$8e$ 	&	1/4  &		1/4  &		0.0074 \\
O(2)      	&$8f$ 	&	0  &		0.9504 &		0.0701  \\
O(3)      	& $8e$ 	&	1/4  &		1/4  &		0.1335 \\
O(4)      	&$8f$ 	&	0  &		0.0373 &		0.2151  \\
O(5)      	&$8e$ 	&	1/4  &		3/4  &		0.1455  \\ [0.2em]
\end{tabular}
\end{ruledtabular}
\end{table*}
\endgroup

\noindent The transformation matrix for conversion of the $Bmab$ orthorhombic unit cell axes to the primitive  unit cell axes is
\[
\begin{bmatrix}
    1      & 0    & 0 \\
    0      & 1    & 0 \\
    -1/2  & 0 & -1/2 
\end{bmatrix}
\]

\vfill
\begingroup
%\squeezetable
\begin{table*}[h]
\begin{ruledtabular}
\centering
\caption{\label{tab:tab2}Crystallographic parameters for monoclinic-II $P2_1/a$ \LNO\  obtained from DFT-LDA.} %and
\begin{tabular}{lcccc}%
\multicolumn{5}{l}{$P2_1/a$ (14)}  $a = 5.41327$~\AA,  $b = 5.46233$~\AA,  $c = 14.23984$~\AA \\
\multicolumn{5}{r}{}$\alpha=\gamma=90^\circ$, $\beta=100.957^\circ$
 \\
Atom    &       Wyck.\ Site     & $x$   & $y$   & $z$   \\
\hline
La(1)        	& $4e$	&   	0.5883 		& 	0  	&		0.1354  \\
La(2)         &$4e$   	&	0.1985  		&	0.4891  	&	0.3959  \\
Ni(1)         &$2a$   	&	0 		&		0  	&		0  \\
Ni(2)        	&$4e$   	& 	0.1417          	&	-0.0028 	& 		0.2781 \\
O(1)      	&$4e$ 	&	0.2932  &		0.2136  &		0.0124 \\
O(2)      	&$4e$ 	&	0.0683  &		-0.0445 &		0.1411  \\
O(3)      	& $4e$ 	&	0.3865 &		0.2511  &		0.2696 \\
O(4)      	&$4e$ 	&	0.2174  &		0.0342 &		0.4304  \\
O(5)      	&$4e$ 	&	0.1021&		1/4  &		0.7096  \\ [0.2em]
\end{tabular}
\end{ruledtabular}
\end{table*}
\endgroup

\vfill
\begingroup
%\squeezetable
\begin{table*}[h]
\begin{ruledtabular}
\centering
\caption{\label{tab:tab3}Crystallographic parameters for  orthorhombic $Pcab$ \LNO\  obtained from DFT-LDA.} %and
\begin{tabular}{lcccc}%
\multicolumn{5}{l}{$Pcab$ (61)}  $a = 5.41327$~\AA,  $b = 5.46233$~\AA,  $c = 27.96049$~\AA \\
Atom    &       Wyck.\ Site     & $x$   & $y$   & $z$   \\
\hline
La(1)        	& $8c$	&   	0.0221 		& 	0.0004  	&		0.5677  \\
La(2)         &$8c$   	&	0.0069  		&	0.0111  	&	0.6980  \\
Ni(1)         &$4a$   	&	0 		&		0  	&		0  \\
Ni(2)        	&$8c$   	& 	0.0071          	&	0.0029 	& 		0.8610 \\
O(1)      	&$8c$ 	&	0.2860  &		0.7853  &		-0.0064 \\
O(2)      	&$8c$ 	&	0.0002  &		0.0457 &		-0.0704  \\
O(3)      	& $8c$ 	&	0.2558 &		0.7486  &		0.8654 \\
O(4)      	&$8c$ 	&	0.0083  &		-0.0349 &		0.7848  \\
O(5)      	&$8c$ 	&	0.2581&		0.2490  &		0.8547  \\ [0.2em]
\end{tabular}
\end{ruledtabular}
\end{table*}
\endgroup

\vfill
\begingroup
%\squeezetable
\begin{table*}[h]
\begin{ruledtabular}
\centering
\caption{\label{tab:tab4}Crystallographic parameters for  the primitive cell of the orthorhombic $Bmab$ of \LNO.} %and
\begin{tabular}{lcccc}%
\multicolumn{5}{l}{$P\bar{1}$ (2)}$a = 5.4133$~\AA,  $b = 5.4623$~\AA,  $c = 14.2398$~\AA \\
\multicolumn{5}{r}{}$\alpha=\gamma=90^\circ$, $\beta=100.957^\circ$\\
Atom    &       Wyck.\ Site     & $x$   & $y$   & $z$   \\
\hline
La(1)        	& $2i$	&   	0.0669 		& 	0.0018  	&		0.1337  \\
La(2)         &$2i$   	&	0.5669  		&	0.4982  	&	0.1337  \\
La(3)        	& $2i$	&   	0.1978 		& 	0.0116  	&		0.3955  \\
La(4)         &$2i$   	&	0.3022  		&	0.5116  	&	0.6045  \\

Ni(1)         &$1d$   	&	1/2 		&		0  	&		0  \\
Ni(2)        	&$1c$   	& 	0          	&	1/2 	& 		0 \\
Ni(3)         &$2i$   	&	0.1385 		&		0.5033  	&		0.2769  \\
Ni(4)        	&$2i$   	& 	0.3615          	&	0.0033 	& 		0.7231 \\

O(1)      	&$2i$ 	&	0.7574  &		1/4  &		0.0148 \\
O(2)      	&$2i$ 	&	0.2574  &		1/4 &		0.0148  \\

O(3)      	& $2i$ 	&	0.0701 &		0.5496  &		0.1403 \\
O(4)      	&$2i$ 	&	0.4299  &		0.0496 &		0.8597  \\

O(5)      	&$2i$ 	&	0.3835  &		1/4  &		0.2670 \\
O(6)      	&$2i$ 	&	0.8835  &		1/4 &		0.26704  \\

O(7)      	& $2i$ 	&	0.2151 &		0.4628  &		0.4301 \\
O(8)      	&$2i$ 	&	0.7151  &		0.0373 &		0.4301  \\

O(9)      	&$2i$ 	&	0.1045&		1/4  &		0.7091  \\
O(10)      	&$2i$ 	&	0.3955&		3/4  &		0.2909  \\ [0.2em]
\end{tabular}
\end{ruledtabular}
\end{table*}
\endgroup

\clearpage

\renewcommand{\figurename}{{Supplementary Figure}}
\setcounter{figure}{0}

\vfill
\begin{figure}[t]
\centering
\includegraphics[width=0.7\columnwidth]{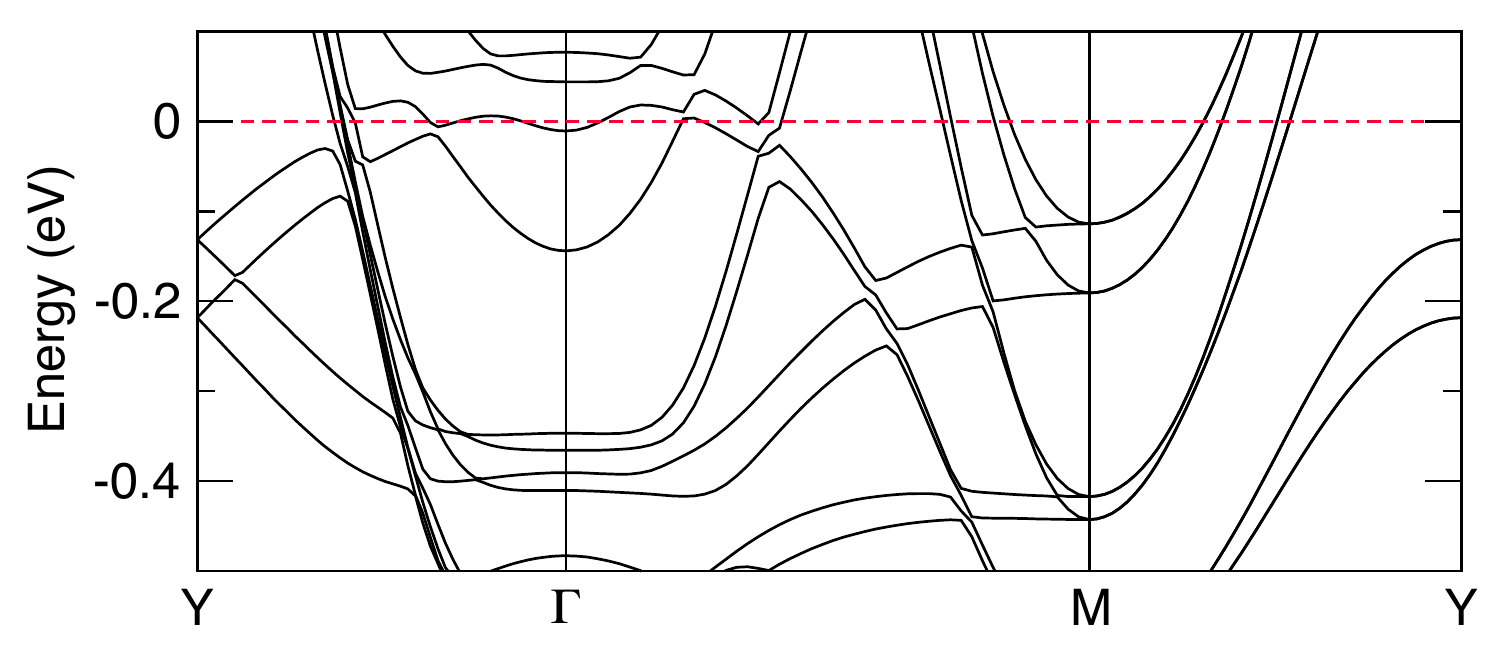}\vspace{-7pt}
\caption{Electronic band structure of the monoclinic-I $P2_1/a$ high-temperature phase of  \LNO.}
\label{figS1}
\end{figure}

\end{document}